\documentclass[showpacs,amsmath,amssymb,twocolumn,prl, floatfix]{revtex4}
\usepackage[dvips]{graphicx}

\def\a{\gamma}
\def\Tr{\operatorname{Tr}}

\begin{document}

\title{Multifractality and intermediate statistics in quantum maps}

\author{J. Martin, O. Giraud and B. Georgeot}
\affiliation{Laboratoire de Physique Th\'eorique, Universit\'e
de Toulouse, CNRS, 31062 Toulouse France}

\date{\today}

\begin{abstract}
We study multifractal properties of wave functions
for a one-parameter family of quantum maps displaying the whole range of
spectral statistics intermediate between integrable and chaotic statistics.
We perform extensive numerical computations and provide analytical
arguments showing that the generalized fractal dimensions are
directly related to the parameter of the underlying classical map,
and thus to other properties such as spectral statistics. Our results
could be relevant for Anderson and quantum Hall transitions, where
wave functions also show multifractality.
\end{abstract}
\maketitle

%%%%%%%%%%%%%%%%%%%%%%%%%%%%%%%%%%%%%%%%%%%%%%%%%%%%%%%%%%%%%%%%%%%%%

%\section*{Introduction}

Statistical properties 
 of wave functions and energy levels 
of quantum
systems  have been an important topic of research in the past decades
with many applications to physical systems. 
It is now well-known
that energy levels of e.g.\ quantum systems whose
classical limit is chaotic, or disordered systems when eigenstates
are extended, follow Random
Matrix Theory (RMT). In this case, wave functions are typically ergodic and
level statistics show level repulsion at short distances.
Conversely, systems whose classical limit is integrable, or
disordered systems in a regime of Anderson localization, show Poisson
statistics of energy levels (without level repulsion), and wave
functions are typically localized in phase space \cite{haake}.

It has been realized only later that another universality class
exists which is
intermediate between the latter two.  It can be observed in
disordered systems at the Anderson transition \cite{braun}, or in
certain systems whose classical limit is pseudo-integrable
\cite{gerland}. In this case, 
level statistics follow specific laws called semi-Poisson
statistics, and wave functions generally
show multifractal properties.  This multifractal
behavior has been extensively studied in the case of the Anderson
transition \cite{Mirlin,EversMirlin,MirlinFyodorov}, 
and has been also seen in quantum Hall transitions \cite{huckestein}.

Recently, a simple model for intermediate systems 
was introduced which corresponds
to a quantization of a certain interval-exchange map \cite{giraud}.
The model, although very simple, can display the
whole range of semi-Poisson statistics when a parameter is changed.
Moreover, a certain randomization of this
system  was shown to yield a new model of Random Matrices with
intermediate statistics \cite{bogomolny}.

Here, we examine multifractal properties of eigenfunctions for the
Random Matrix model corresponding to intermediate quantum maps. We
compute the inverse participation ratios (IPR), fractal
dimensions, and singularity spectra in a variety of regimes with
different numerical methods. Using extensive numerical studies and
analytical arguments, we show that the parameter of
the model can be related 
to the fractal dimensions of the eigenfunctions, as well
as to the spectral statistics. This Random Matrix model is known to
span the whole range of semi-Poisson statistics
for both short-range and long-range
statistics.  Thus our results indicate with some generality the existence of
a link between the statistics of eigenvalues and the
multifractal properties of the eigenfunctions. 
%We also test several
%hypotheses which have been linked to the Anderson transition using
%the nonlinear sigma-model, to assess if they are a generic feature
%of intermediate systems.  
%Our results shed some light on the
%relationship between multifractality of quantum systems
%and their other properties.
%Let us start with the classical map defined on the 2-torus by
%\begin{equation}
%\Phi_{\a}: \begin{pmatrix} p \\ q \end{pmatrix} \mapsto
%\begin{pmatrix}p + \a \\ q + 2(p+\a) \end{pmatrix}\mathrm{(mod} \;\mathrm{1)}.
%\end{equation}
%where $(p,q)$, the coordinates in phase space, are the conjugated
%momentum (action) and angle variables.

Let us start with the classical map defined on the 2-torus by
$\Phi_{\a}: \bar{p} = p + \a \;\mathrm{(mod} \;\mathrm{1)}\;;\;
\bar{q} = q+ 2\bar{p} \;\mathrm{(mod} \;\mathrm{1)}$, where
$(p,q)$, the coordinates in phase space, are the conjugated action
and angle variables and the bars denote the resulting variables
after one iteration of the map. The quantization of this map yields
a unitary evolution operator acting on a Hilbert space of dimension
$N=1/(2\pi\hbar)$ which can be expressed in momentum space by the $N
\times N$ matrix~\cite{giraud,bogomolny}
\begin{equation}\label{ISRM}
    U_{pp'}=\frac{e^{i\phi_p} }{N}\frac{1-e^{2i\pi N
\a}}{1-e^{2i\pi (p-p'+N\a)/N}},
\end{equation}
with $\phi_p=-2\pi p^2/N$. From this quantized map one can
construct an ensemble of random matrices, taking $\phi_p$ as random
variables uniformly distributed in $[0, 2\pi[$ \cite{bogomolny}. 
The statistical
properties of the pseudo-spectrum (the set of eigenphases) of $U$
strongly depend on the value of the parameter $\a$. On the one
hand, for generic irrational $\a$, the spectral statistics of $U$
are expected to follow those of the Circular Unitary Ensemble (CUE)
of RMT if the $\phi_p$ are independent (non-symmetric case),
or the Circular Orthogonal
Ensemble (COE) if one imposes a symmetry $\phi_{N-p}=\phi_p$. On the
other hand, for rational $\a=a/b$, a variety of different behaviors
are observed \cite{giraud}.  It was shown in \cite{bogomolny} that
for $aN=\pm 1\;\mathrm{mod}\;b$ the spectral statistics is of
semi-Poisson type. In particular the nearest-neighbor spacing
distribution is given by $P_{\beta}(s)=A_\beta s^\beta
e^{-(\beta+1)s}$ with parameter $\beta =b-1$ in the non-symmetric
case ($\beta =b/2 -1$ in the symmetric case). For $aN\neq \pm
1\;\mathrm{mod}\;b$, $P(s)$ is still of intermediate type but with
more complicated formulas \cite{private}. Finally when $\a$ is an
integer the eigenphases are equally spaced and the spectrum is
totally rigid. Thus the set of quantum maps $U$ with rational $\a$
gives a random
matrix ensemble with intermediate statistics (ISRM) whose spectral
statistics correspond to natural intermediate distributions between
Poisson and RMT, controlled by the value of $\a$.

Multifractality properties of wave functions are described by a
whole set of generalized fractal dimensions $D_q$. For a vector
$|\psi\rangle=\sum_{i=1}^{N}\psi_i|i\rangle$ in an $N$-dimensional
Hilbert space, the multifractal exponents $D_q$ are defined through
the scaling of the moments
\begin{equation}
\label{dimfrac}
\sum_{i=1}^N |\psi_i|^{2q} \propto N^{-D_q(q-1)}.
\end{equation}
The fractal dimension for $q=0$ corresponds to the dimension of the
support of the measure, here $D_0=1$. The fractal exponent $D_2$
describes the large-size behavior of the IPR
$\xi=1/\sum_{i=1}^N|\psi_i|^4$, which measures the extension of the
state $|\psi\rangle$ over the basis vectors. The multifractal
exponents describe the behavior of the partition function
\begin{equation}
Z(q,L)\equiv\sum_{k=1}^{N_b}\mu_k(L)^q\propto L^{\tau_q},\ \ \ \ \
\ \tau_q \equiv D_q(q-1), \label{Zqtauq}
\end{equation}
where the vector $|\psi\rangle$ is divided into $N_b=N/L$ boxes
$B_k$ of size $L$, and $\mu_k(L)=\sum_{i\in B_k}|\psi_i|^2$, $1\leq
k\leq N_b$. The multifractal properties are alternatively
characterized by the singularity spectrum $f(\alpha)$, which is the
fractal dimension of the set of points whose singularity exponent is
$\alpha$. It is related to the function $\tau_q$ by a Legendre
transform. Introducing the normalized measures
$\mu_k(q,L)=\mu_k(L)^q/\sum_i\mu_i(L)^q$, the singularity exponent
and the associated fractal dimension can respectively be obtained
by~\cite{Chhabra}
\begin{equation}
\label{falphaq}
\begin{aligned}
\alpha (q)&=\frac{d\tau_q}{dq} =\lim_{L/N\to 0}\frac{\sum_i\mu_i(q,L)\log \mu_i(L)}{\log
(L/N)},\\
f(\alpha (q))&= q\,\alpha (q)-\tau_q =\lim_{L/N\to 0}\frac{\sum_i\mu_i(q,L)\log\mu_i(q,L)}{\log (L/N)}.
\end{aligned}
\end{equation}
%\begin{equation}
%\label{alphaq}
%\alpha (q)=\lim_{L\to 0}\frac{\sum_i\mu_i(q,L)\log \mu_i(L)}{\log L}
%\end{equation}
%and
%\begin{equation}
%\label{falphaq}
%f(\alpha (q))=\lim_{L\to 0}\frac{\sum_i\mu_i(q,L)\log\mu_i(q,L)}{\log L}.
%\end{equation}
%The multifractal exponent $D_q$ is expressed in terms of the
%singularity spectrum as $D_q=(f(\alpha (q))-q\,\alpha (q))/(1-q)$.

\begin{figure}
\begin{center}
\includegraphics[width=.95\linewidth]{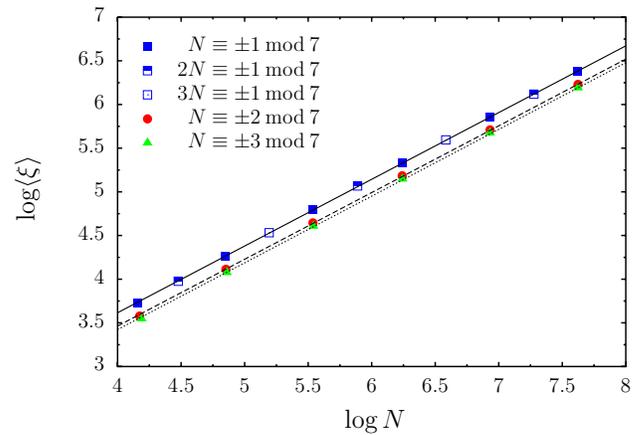}
\end{center}
\caption{(Color online) Mean IPR of eigenvectors of (\ref{ISRM})
as a function of the vector size
$N$ for $\a=a/b$ with $b=7$ and $a=1$ (filled symbols), $a=2$
(half-filled squares), $a=3$ (empty squares). Straight lines
correspond to the best linear fits. Logarithm is natural.}
\label{Nmodloglog}
\end{figure}

Let us consider an ensemble of matrices of type \eqref{ISRM}
with rational $\a=a/b$, in the non-symmetric case where all $\phi_p$
are independent. The mean IPR for eigenvectors of these matrices
in $p$ representation for different
values of $\a$ with denominator $b=7$ is displayed in
Fig.~\ref{Nmodloglog} as a function of the matrix size. The IPR
scales as $N^{D_2}$ provided data corresponding to different values
of $aN\;\mathrm{mod}\;b$ be treated separately. Indeed, when
different matrix sizes are grouped into families, the results yield a
linear behavior of $\log\langle \xi\rangle$ as a function of $\log
N$, with the same slope $D_2$ for each family. More generally we
observed that the fractal exponents $D_q$ are well defined if data
are organized into families, and that they only depend on the
denominator $b$ of $\a$.

We now proceed to compute the fractal exponents $D_q$. A few of
these exponents have already been computed in \cite{bogomolny, georgeot} for
the case $aN\equiv\pm 1\;\mathrm{mod}\;b$. Here our aim is to
characterize $D_q$ as a function of $q$. The quantities $D_q$ and
$f(\alpha)$ are known to be difficult to compute numerically,
especially for large $q$ or $\alpha$. In this work, we resorted to
several different methods as a consistency check. We first opted for
the usual method of moments. We computed average values of the
moments \eqref{dimfrac} for different system sizes $N$ ranging from
$\sim2000$ up to $\sim12000$ to get rid as much as possible of
finite size effects \cite{remark}. The fractal dimensions are
extracted from the slopes of the graphs of $\log \langle\sum_i
|\psi_i|^{2q}\rangle$ versus $\log N$. Here the average is taken
over all eigenvectors and random realizations of $U$ (from $200$
realizations for $N\sim 2000$ to $1$ for $N\sim 12000$).
%$200$ for $N\sim 2000$, $35$ for $N\sim 4000$, $7$ for $N\sim 6000$,
%$1$ for $N\sim 8000$ and $N\sim 12000$.
We also opted for the so-called canonical method \cite{Chhabra}
allowing to determine the $f(\alpha)$ spectrum directly from
Eq.~\eqref{falphaq}. For this method, the numerical computations
were done on a single realization of size $N\sim 13000$, and $20$
box sizes ranging from $L=10$ to $L=0.1 N$ (again different families
of box sizes were treated separately). We also considered other
approaches, such as the box counting method based on
Eq.~\eqref{Zqtauq}; they all give results intermediate between the
two previous methods.
\begin{figure}
\begin{center}
\includegraphics[width=.95\linewidth]{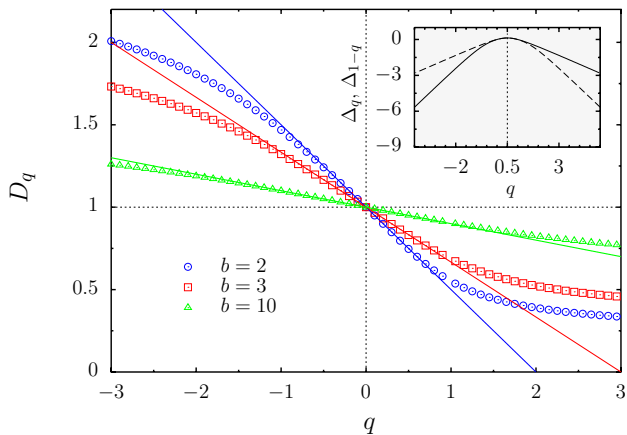}
\end{center}
\caption{(Color online) Fractal dimension $D_q$ computed with the
canonical method for $\a=1/b$ with $b=2$ (blue circles), $b=3$ (red
squares), $b=10$ (green triangles). Solid lines show the linear
approximation $D_q=1-q/b$. Inset: anomalous exponent $\Delta_q$ as a
function of $q$ (solid line), together with its symmetry with
respect to $q=1/2$ (dashed line) for $b=2$.} \label{DDq}
\end{figure}

\begin{figure}
\begin{center}
\includegraphics[width=.95\linewidth]{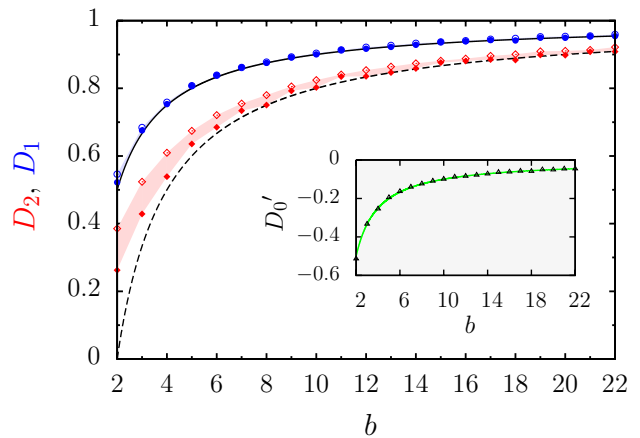}
\end{center}
\caption{(Color online) Information dimension $D_1$ (blue circles)
and correlation dimension $D_2$ (red diamonds) as a function of the
denominator of $\a=1/b$. Full (empty) symbols are numerical values
obtained by the method of moments (the canonical method). Solid and
dashed lines are the theoretical curves $1-1/b$ and $1-2/b$ for
$D_1$ and $D_2$ respectively. The values obtained by e.g.\ the box
counting method lie in the shaded domain in between. Inset: slope of
$D_q$ at the origin $q=0$ (triangles) and curve $-1/b$ (solid
line).} \label{D1D2}
\end{figure}

The results for $D_q$ are displayed in
Fig.~\ref{DDq}.  For increasing $b$, the curve for
$D_q$ tends slowly to the limiting curve where $D_q=1$ for all $q$,
which corresponds to non-fractal wave functions. This is in
agreement with the fact that $\a$ tends 
for $b\to \infty$ to an
irrational number for which the system is expected to follow RMT.
Figure \ref{DDq} shows that $D_q$ is roughly linear in a
relatively large interval around $q=0$, and tends to limiting values
$D_{\pm \infty}$ for large $|q|$.
The slope of $D_q$ at $q=0$ is
displayed in the inset of Fig.~\ref{D1D2} as a function of $b$. We found that the value
of this slope is very accurately given by $-1/b$. Since $D_0=1$, the
first-order expansion of $D_q$ around $q=0$ reads
\begin{equation}
\label{DQ} D_q \approx 1-\frac{q}{b}.
\end{equation}
This expansion turns out to be valid in a quite large interval
of $q$, whose size increases with $b$.  As an example, the
numerical values of $D_1$ and $D_2$ together with the linear
approximation \eqref{DQ} are shown in Fig.~\ref{D1D2}. When $b$ is
increased, Eq.~\eqref{DQ} is verified with higher and higher
accuracy. Of course Eq.~\eqref{DQ} breaks
down for large $|q|$ since $D_q$ is bounded. 
%Remarkably enough, our formula has a functional form similar
%to the result obtained in \cite{efetov} for the particular case of
%two-dimensional 
%disordered systems for samples smaller than the localization length.

To get an understanding of why formula \eqref{DQ} holds, we note that 
general arguments for critical systems predict that the
multifractal properties of the eigenstates for $q=2$ 
are linked to the spectral statistics
through a relation between the correlation dimension $D_2$ and the
level compressibility $\chi$~\cite{Chalker},
\begin{equation}
\label{chalker} \chi=\frac{1}{2}\left(1-\frac{D_2}{D_0}\right).
\end{equation}
 Numerical results for the power-law random banded matrix
ensemble \cite{EversMirlin} have revealed that this relation is
extremely well verified in the regime of weak multifractality (large
bands): in this model, the fractal dimension evolves linearly with
respect to $q$ as $D_q=1-\kappa q$, where $\kappa$ is inversely proportional
to the
width of the central band (and in this particular case can be also
related to the level compressibility). However, for smaller bands
Eq.~\eqref{chalker} was clearly violated. Suppose
Eq.~\eqref{chalker} holds in our case. For ISRM (Eq.~\eqref{ISRM}),
the level compressibility can be estimated analytically. It is given
by the value of the two-point correlation form factor
$K_2(\tau)=|\Tr U^n|^2/N$, with $\tau=n/N$, for $n/N\to 0$.
Following \cite{giraud}, we note that in the semiclassical limit
$N\to\infty$ and fixed $n$, the trace $\Tr U^n$ is asymptotically
equal to $\Tr V_n$, where $V_n$ is the quantization of the $n$th iterate
of the classical map, and
\begin{equation}
\Tr V_n = \frac{1}{N} \sum_{p=0}^{N-1} \exp( i n \phi_p)
\sum_{k=0}^{N-1} \exp(2 i \pi n \a k).
\end{equation}
The modulus squared of the first sum yields $\approx N$ when all $\phi_p$
are random. 
%(nonsymmetric case) and $2N$ when only a half of 
%the $\phi_p$ is independent (symmetric case).
The second sum is a geometric sum: for $\a=a/b$, it is equal to $N$
if $n$ is divisible by $b$ and to $O(1)$ otherwise. Thus,
$K_2(n/N)\sim 1$ provided $n$ is divisible by $b$, $K_2(n/N)\sim 0$
in all other cases. The level compressibility is then given by the
time averaged form factor
\begin{equation}\label{K0}
\chi=\overline{K_2(0)} \equiv \lim_{n\to\infty} \lim_{N\to\infty}
\frac1n \sum_{n'=1}^n K_2(n'/N) \approx \frac{1}{b}.
\end{equation}
%in the non-symmetric case. and $1/(2b)$ in the symmetric one.
Inserting this value of $\chi$ into Eq.~\eqref{chalker} we get
$D_2\approx 1-2/b$, which corresponds to Eq.~\eqref{DQ} for $q=2$.
A simple linear interpolation between this value for $D_2$ and
$D_0=1$ yields Eq.~(\ref{DQ}).  We note that for small $b$ (strong
multifractality) Eq.~\eqref{chalker} breaks down but 
Eq.~(\ref{DQ}) is still valid for smaller $q$ values.

Before moving to the study of the singularity spectrum,
we briefly discuss  symmetry properties of $D_q$.
 It was suggested 
in \cite{MirlinFyodorov} that
the anomalous exponents $\Delta_q$, defined by
$\Delta_q\equiv (D_q-1)(q-1)$, approximately follow the symmetry
relation
$\Delta_q=\Delta_{1-q}$.
This was shown to hold for the Anderson model with good accuracy
over a large interval of $q$ values. It is not the case in our
system. As an example, 
the inset in Fig.~\ref{DDq} gives $\Delta_q$ and
$\Delta_{1-q}$ for $\a=1/2$. For values of $q$ where the exponents
$D_q$ have a linear behavior the symmetry relation holds, as it should
since any linear $D_q$ necessarily fulfills it. Outside
the linear regime, the relation is not verified anymore.

\begin{figure}
\begin{center}
\includegraphics[width=.95\linewidth]{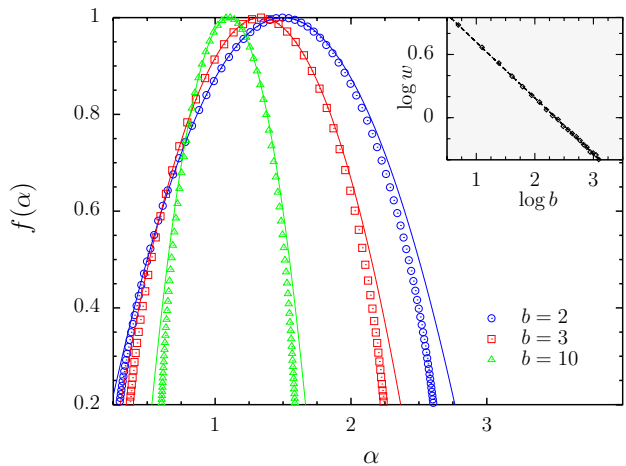}
\end{center}
\caption{(Color online) Singularity spectrum $f(\alpha)$ for
$\a=1/b$ with $b=2$ (blue circles), $b=3$ (red squares), $b=10$
(green triangles). Solid lines show the parabola~\eqref{falpha}.
Inset: singularity spectrum width
$w=\alpha_{\mathrm{max}}-\alpha_{\mathrm{min}}$ as a function of
$b$. The dashed line shows the best linear fit $\log w = -0.530\log
b+1.247$. Logarithm is natural.} \label{fa}
\end{figure}

We now turn to the singularity spectrum $f(\alpha)$. For $\a=a/b$,
the expression obtained using Eq.~\eqref{DQ} is
\begin{equation}
\label{falpha}
f(\alpha)\approx 1-\frac{b}{4}\left(\alpha-1-\frac{1}{b}\right)^2.
\end{equation}
It reaches its maximum at $\alpha(q=0)=1+1/b$. Since  Eq.~\eqref{DQ}
is valid around $q=0$, we expect  Eq.~\eqref{falpha} to be accurate
around $\alpha(0)$. Figure \ref{fa} shows the singularity spectrum,
numerically computed using Eq.~\eqref{falphaq}, together with the
theoretical estimate Eq.~\eqref{falpha}. The data displayed show
that \eqref{falpha} approximates the singularity spectrum with good
accuracy over a large interval of values of $\alpha$ around
$\alpha(0)$. As $b \rightarrow \infty$, the curve for $f(\alpha)$
gets closer and closer to a single point at $\alpha=1$, once again
corresponding to the non-fractal case of RMT.

We previously showed that
the theory (\ref{DQ}) accurately describes the behavior of the
moments in Eq.~\eqref{dimfrac} for small $q$.  For large values of
$q$, this is bound to break down since $D_q$ converges to a finite
asymptotic value for $q \rightarrow \pm\infty$. Similarly
$f(\alpha)$ should have vertical asymptotes at some limiting values
$\alpha_{\mathrm{max}}$ and $\alpha_{\mathrm{min}}$, while
\eqref{falpha} is the equation of a parabola. However numerical data
are not far from the theory \eqref{falpha}, and some of the features
of $f(\alpha)$ are well captured by this estimate. For example, the
inset of Fig.~\ref{fa} shows that the width
$w=\alpha_{\mathrm{max}}-\alpha_{\mathrm{min}}$ of the singularity
spectrum scales as $\sim 1/b^{0.53}$ (best fit). This is close to the scaling
$1/\sqrt{b}$ of the difference between the two intersections of the parabola
\eqref{falpha} with a straight line.
%(the difference between the two roots of $f(\alpha)=c$ is
%$4\sqrt{(1+c)/b}$).
%A simple
%interpolation between the two behaviors is to use (\ref{falpha}) up
%to some value $f(\alpha)=c<1$ and then replace $f(\alpha)$ with
%vertical lines for $f(\alpha)<c$. The value of $c$ should go to zero
%as $b\rightarrow \infty$.  In this simple approximation,
%$w=\alpha_{\mathrm{max}}-\alpha_{\mathrm{min}}$ is given by the
%difference between
% the two roots of $f(\alpha)=c$ with $f(\alpha)$ given by (\ref{falpha}).
%This predicts $\alpha_{\mathrm{max}}-\alpha_{\mathrm{min}} \sim
%1/\sqrt{b}$.

We finish by noting that in the symmetric case ($\phi_{N-p}=\phi_p$
in Eq.~\eqref{ISRM}) we have performed similar computations, getting
very similar data. In particular, Eqs.~\eqref{DQ} and \eqref{falpha}
are valid in this case as well.

In conclusion, we have studied multifractal properties of
eigenfunctions for intermediate quantum maps. Although data
corresponding to system sizes $N$ with different values of
$aN\;\mathrm{mod}\;b$ should be treated separately, they give the
same value for $D_q$ and $f(\alpha)$. Our results show that for an
interval of $q$ values whose size increases with $b$, the fractal
exponents can be related explicitly to the parameter $\a=a/b$ of the
map through Eq.~\eqref{DQ}, and thus to spectral statistics. A
similar result holds for the singularity spectrum through
Eq.~\eqref{falpha}. 
Thus in such a system, fractal exponents and
singularity spectrum are related to the spectral properties 
 over a wide range of fractal
dimensions. Interestingly enough, our relation is still valid
for small $q$ even when Eq.~\eqref{chalker} for $q=2$ does not hold.
As our system corresponds to a Random Matrix model
covering the whole range of semi-Poisson statistics, both at short-range
and long-range, we can expect our results to display some generality.
It will be interesting to study if similar results apply
to other intermediate systems, and other physical systems where
wave functions are multifractal, such as condensed-matter
systems at the Anderson or quantum Hall transitions.

We thank E.~B.~Bogomolny and K.~Frahm for helpful discussions,
CalMiP and IDRIS
for access to their supercomputers, and
the French ANR
(project INFOSYSQQ) and the IST-FET program of the EC
(project EUROSQIP) for funding.


\begin{thebibliography}{99}
\bibitem{remark} As $U$ is not defined for $aN= 0\;\mathrm{mod}\;b$, dimensions $N$ are not necessarily the same for all $b$.
\end{thebibliography}
\end{document}